\documentclass[twocolumn,aps,prl,draft,showpacs]{revtex4}
\usepackage{graphicx}
\usepackage{bm}
\begin{document}
\title{Two-dimensional turbulence of dilute polymer solutions}
\author{Guido Boffetta$^1$, Antonio Celani$^2$, and Stefano Musacchio$^1$}
\affiliation{$^1$ Dipartimento di Fisica Generale and INFM, 
Universit\`a degli Studi di Torino, V. Pietro Giuria 1, 10125, Torino, Italy \\
$^2$ CNRS, INLN, 1361 Route des Lucioles, 06560 Valbonne, 
France}
\date{\today}
\begin{abstract}
We investigate theoretically and numerically the effect of polymer 
additives on two-dimensional turbulence by means of a viscoelastic model. 
We provide compelling evidence that at vanishingly small concentrations, 
such that the polymers are passively transported, the probability 
distribution of polymer elongation has a power law tail: its slope is 
related to the statistics of finite-time Lyapunov exponents of the flow, 
in quantitative agreement with theoretical predictions. 
We show that at finite concentrations
and sufficiently large elasticity the polymers react on the flow
with manifold consequences: velocity fluctuations 
are drastically depleted, as observed in soap film experiments; 
the velocity statistics becomes strongly intermittent; the distribution
of finite-time Lyapunov exponents shifts to lower values,
signalling the reduction of Lagrangian chaos.
\end{abstract}
\maketitle 
Since the discovery of the conspicuous drag reduction obtained by dissolving
minute amounts of long chain molecules in a liquid, turbulence of 
dilute polymer solutions
has attracted a lot of attention in view of its
industrial applications (see, e.g. Refs.~\cite{L69,NH95,SW00}).
The fluid mechanics of polymer solutions is appropriately described 
by viscoelastic models that are able to reproduce the rheological behavior
and many other experimental observations \cite{BCAH87}. For example,
it has been shown by Sureshkumar et al.
that the drag reduction effect can be captured by numerical simulations of 
the channel flow of viscoelastic fluids \cite{SBH97}. 
Although the parameters used in those simulations do not
match the experimental ones, the qualitative agreement
is remarkable, and all the hallmarks of the turbulent flow 
of polymer solutions are recovered in numerical experiments.

Following this premise, it is natural to ask whether a two-dimensional
viscoelastic model can reproduce the recent 
results by Amarouchene and Kellay \cite{AK02} 
showing that the turbulent flow of soap films is spectacularly affected by
polymer additives (see also Refs.~\cite{CBG01,PB99}). 
Here we show that this is indeed the case, and that 
the suppression of large-scale velocity fluctuations
observed experimentally has a 
simple theoretical explanation. 
However, the influence of polymers is not limited to the depletion of 
mean square velocity, which is a genuinely two-dimensional
effect. In the viscoelastic case we observe
a strong intermittency, with exponential tails of the 
velocity probability density. 
As for the Lagrangian statistics, we show that the values of
finite-time Lyapunov exponents lower significantly upon polymer
addition, which therefore reduces the chaoticity of the flow.  
These effects are expected to be independent of the space dimensionality,
and thus relevant to three-dimensional turbulence as well.

We also investigate the limit of vanishingly small polymer concentrations, 
in which the polymer molecules have no influence on the advecting flow.
In this case the velocity field evolves according the two-dimensional 
Navier-Stokes equation with friction, and is therefore smooth 
at scales smaller than the injection lengthscale \cite{KM80,NOAG00}.
For passive polymers, space dimensionality plays only a minor role,
and our system is an instance of a generic
random smooth flow to which
the theory of passive polymers developed by Chertkov \cite{C00} 
and Balkovsky et al. \cite{BFL00,BFL01} applies.
We check this theory against our numerical results, 
and find an excellent quantitative agreement.

To describe the dynamics of a dilute polymer solution we 
adopt the linear viscoelastic model (Oldroyd-B) 
\begin{equation}
\partial_t {\bm u} + ({\bm u}\cdot{\bm \nabla}) {\bm u}
=-{\bm \nabla p} + \nu {\Delta} {\bm u} + \frac{2 \eta\,\nu}{\tau} 
{\bm \nabla}\cdot{\bm \sigma} -\alpha {\bm u}
+ {\bm f}
\label{eq:1}
\end{equation}
\begin{equation}
\partial_t {\bm \sigma} + ({\bm u}\cdot{\bm \nabla}) {\bm \sigma}
= ({\bm \nabla \bm u})^T \cdot {\bm \sigma} + {\bm \sigma} \cdot 
({\bm \nabla \bm u})
-2\frac{({\bm \sigma}-{\bm 1})}{\tau}\;.
\label{eq:2}
\end{equation}
The velocity field ${\bm u}$ is incompressible,
the symmetric matrix  ${\bm \sigma}$ is
the conformation tensor of polymer 
molecules, and its trace $\textrm{tr}{\bm \sigma}$ is a measure of 
their elongation \cite{note}.  
The parameter $\tau$ is the (slowest) polymer relaxation time.
The energy source
${\bm f}$ is a large-scale random, zero-mean, 
statistically homogeneous and isotropic, solenoidal vector field.
The pressure term $-{\bm \nabla}p$ ensures incompressibility of the 
velocity field. The matrix of velocity gradients is defined as 
$({\bm \nabla \bm u})_{ij}=\partial_i u_j$ and ${\bm 1}$ is the unit tensor.
The solvent viscosity is denoted by $\nu$ and
$\eta$ is the zero-shear contribution of polymers to the total solution
viscosity $\nu_{t}=\nu(1+\eta)$. 
The dissipative term $-\alpha {\bm u}$ models the 
mechanical friction between the soap film and the surrounding air \cite{RW00},
and plays a prominent role in the energy budget
of Newtonian two-dimensional turbulence \cite{BCMV02}. 
It should be remarked that a model that describes more accurately the 
polymer dynamics is the FENE-P model, which accounts for the nonlinear
character of polymer elasticity, culminating in a finite 
molecular extensibility \cite{BCAH87}. Although here we limit 
ourselves to the linear case because it allows a simpler theoretical 
treatment, our conclusions apply to the nonlinear case as well,
provided that the maximal polymer elongation is very large compared
to the equilibrium length. 

{\em Passive polymers}. The effect of polymer concentration $n$ is 
included in Eq.~(\ref{eq:1}) through the parameter $\eta \propto n$.
In the limit $\eta = 0$ polymers are passively transported and 
stretched by Newtonian two-dimensional turbulence.
The flow is driven at the largest scales and develops an enstrophy
cascade towards the small scales, while the inverse energy flux is
immediately halted by friction. The ensuing velocity field is 
therefore everywhere smooth. We briefly recall that 
according to Refs.\cite{C00,BFL00,BFL01} the statistics of stretched 
polymers in random smooth flows is expected to depend critically 
on the value of the Weissenberg number, here defined as $Wi= \lambda_N \tau$,
where $\lambda_N$ is the maximum Lyapunov exponent of the Newtonian flow.
At $Wi < 1$ the polymer molecules spend most of the time 
in a coiled state, and stretch occasionally by a considerable amount.
\begin{figure}[b]
\centerline{\hspace{-0.5cm}
\includegraphics[draft=false, scale=0.7]{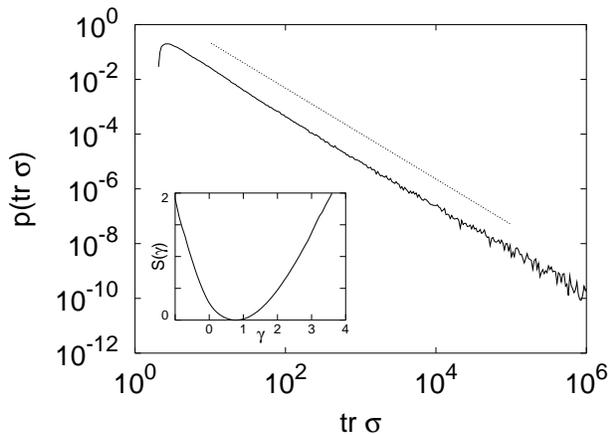}} 
\caption{Power law tail of the probability density 
function of polymer square elongation,
in the passive case $\eta=0$.
The Weissenberg number is $Wi=0.4$,  
quite below the coil-stretch transition.
The power law $(\textrm{tr}{\bm \sigma})^{-1-q}$ with the theoretical value 
$q=0.66$ (obtained from the relation $L_{2q}=2q/\tau$) 
is drawn for comparison. In the inset, the corresponding
Cram\'er functions $S(\gamma)$. Its minimum is $S(\lambda_N)=0$,
with $\lambda_N \simeq 0.8$. 
Data have been obtained by direct numerical simulation of 
the equation of the conformation tensor (\protect\ref{eq:2}) 
by a Lagrangian method (see, e.g. \protect\cite{HLKL98} and 
references therein), while 
Eq.~(\protect\ref{eq:1}) has been solved in a doubly periodic box 
by a pseudospectral code at resolution $256^2$. 
The velocity field is driven by a 
 Gaussian, homogeneous, isotropic
 $\delta$-correlated in time forcing, with correlation
length $L \approx 4$. The Reynolds number is 
$Re=u_{rms}L/\nu\approx 4000 $.}
\label{fig:1} 
\end{figure}
The theory predicts a power law tail for 
the probability density function of $\textrm{tr}{\bm \sigma}$, i.e.
the square polymer elongation
\begin{equation}
p(\textrm{tr} {\bm \sigma}) \sim (\textrm{tr} {\bm \sigma})^{-1-q}
\qquad \mbox{for}\qquad \textrm{tr}{\bm \sigma} \gg \textrm{tr}{\bm 1} \;.
\label{eq:3}
\end{equation}    
The exponent $q$ is related to the probability
 of finite-time Lyapunov exponents
$P(\gamma,t)\propto \exp[-t S(\gamma)]$ via the equation 
$L_{2q}=2q/\tau$, where  
 $L_{2q}=\max_{\gamma}[2q\gamma-S(\gamma)]$ 
is the generalized Lyapunov exponent of order $2q$, 
and $S(\gamma)$ is the Cram\'er rate function (see, e.g., Ref.~\cite{CPV93}). 
The convexity of $S(\gamma)$ ensures
the positivity of $q$ for $Wi<1$.
Since the distribution of polymer elongations is not accessible 
experimentally, in order to validate the theory it is necessary
to resort to numerical simulations. 
Eckhardt et al. in Ref.~\cite{eckhardt} have given the first evidence
of  a power law tail for the 
probability distribution function of
polymer elongation in three dimensional shear turbulence.
As shown in Fig.~\ref{fig:1}, in our two-dimensional simulations 
we observe a neat power law as well.
In order to check whether the observed exponent coincides with
the predicted one, we have also 
performed direct numerical simulations of particle 
trajectories, and measured the probability distribution of finite-time
Lyapunov exponents, thereby obtaining the expected $q$. 
The numerical result is in close agreement with theory. 
As the Weissenberg number exceeds unity, the probability distribution 
of the conformation tensor becomes unstationary and all moments
$\langle (\textrm{tr}{\bm \sigma})^n \rangle$ grow exponentially in time.
This ``coil-stretch'' transition signals the breakdown of linear
passive theory.
Accounting for the nonlinear elastic modulus of 
polymer molecules allows to recover a stationary statistics and to 
develop a consistent theory of passive polymers at all Weissenberg numbers
\cite{T02}. In the following we do not pursue that approach, but we rather
focus on a different mechanism that limits polymer elongation:
the feedback of polymers on the advecting flow.

\begin{figure}[b]
\centering
\includegraphics[draft=false, scale=0.36, clip=true]{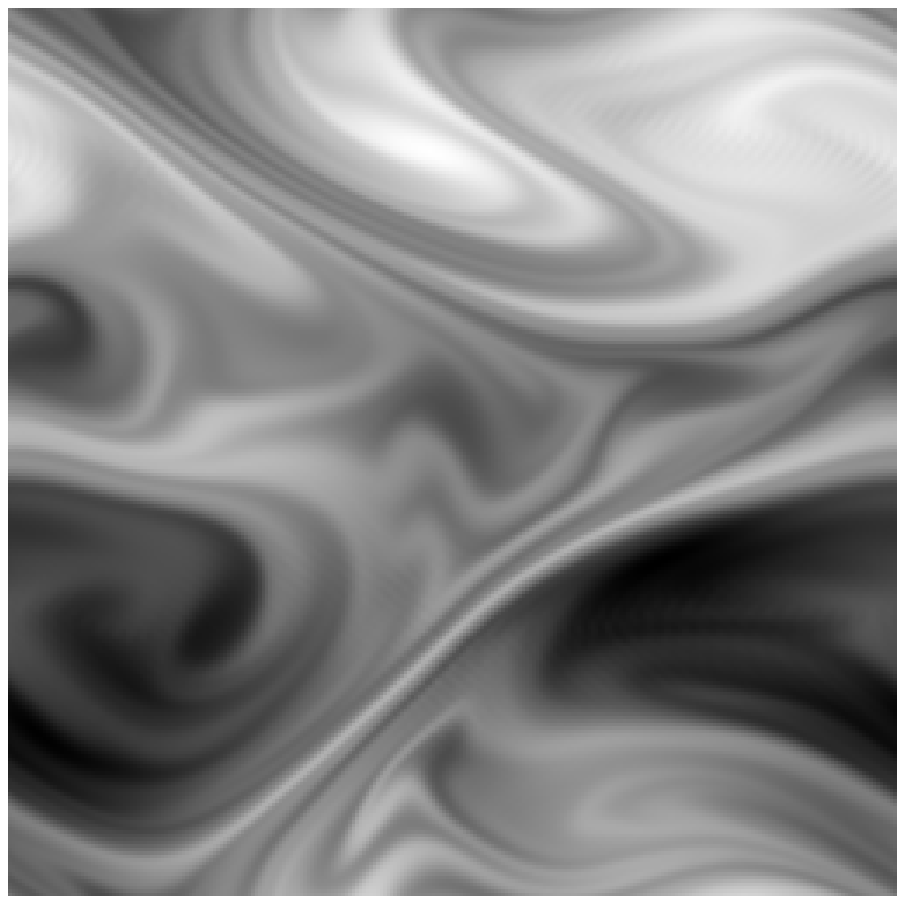}
\hspace{0.2cm}
\includegraphics[draft=false, scale=0.36, clip=true]{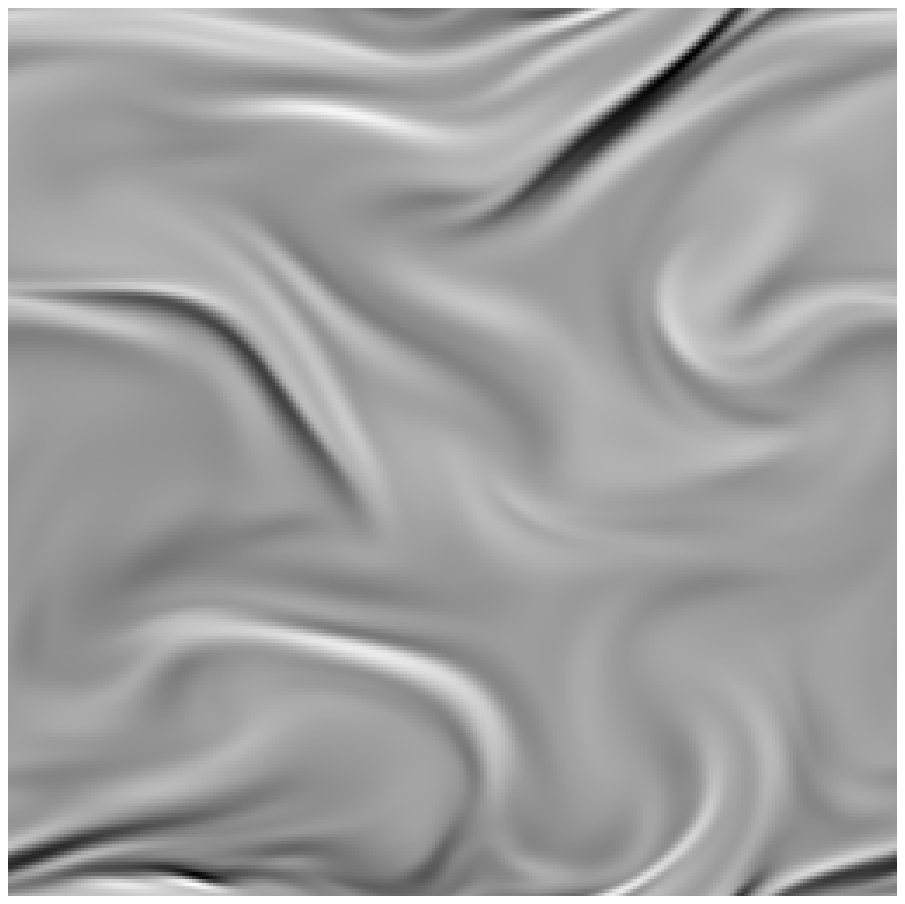}
\caption{ Snapshots of the vorticity field ${\bm \nabla} \times {\bm u}$
in the Newtonian (left) and in the viscoelastic case with strong feedback 
(right).
Notice the suppression of large-scale structures
in the latter case.
The fields are obtained by a fully dealiased pseudospectral simulation of
eqs.~(\ref{eq:1})-(\ref{eq:2}) at resolution $256^2$. 
The viscosity is $\nu=1.5\cdot 10^{-3}$, $\eta=0.2$, the relaxation time is 
$\tau=4$, the energy input is $F=3.5$. 
As customary, an artificial stress-diffusivity term is added 
to eq.(\ref{eq:2}) to prevent numerical instabilities \protect\cite{SB95}. 
The corresponding Schmidt number is $Sc=0.25$.}
\label{fig:2} 
\end{figure}

\begin{figure}[h]
\centerline{\hspace{-0.5cm}
\includegraphics[draft=false,scale=0.7]{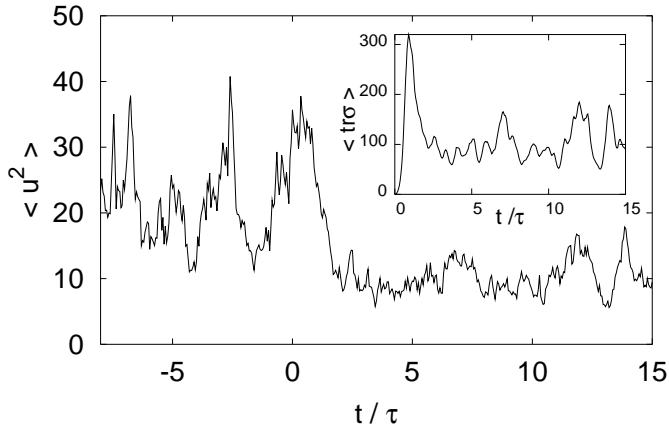} }
\caption{Dilute polymers reduce the level of velocity fluctuations 
$\int |{\bm u}({\bm x},t)|^2\,d{\bm x}$. Polymers are introduced in the 
flow at $t=0$. In the inset, the mean square 
elongation $\int \textrm{tr}{\bm \sigma}({\bm x},t)\, d{\bm x}$ as a function
of time.}
\label{fig:3} 
\end{figure}

{\em Active polymers}. When $\eta > 0$, polymers can affect 
significantly the velocity dynamics, provided that they are sufficiently
elongated -- a condition that is met at $Wi > 1$. 
This strong feedback regime is characterized in two dimensions by a 
suppression of large-scale velocity
fluctuations (see Fig.~\ref{fig:2}), an effect  first observed in soap film 
experiments \cite{AK02}. 
In Fig.~\ref{fig:3} we present the time evolution
of the total kinetic energy of the system, showing that after polymer 
injection a drastic depletion of kinetic energy occurs.
This should be contrasted with the 
three-dimensional case where, on the opposite, velocity fluctuations 
are larger in the viscoelastic case than in the Newtonian one \cite{DCBP02}.

The suppression of velocity fluctuations by polymer additives
in two-dimensional turbulence can be easily explained
in the context of the randomly driven viscoelastic model 
(\ref{eq:1})-(\ref{eq:2}).
Indeed, the average kinetic energy balance in the statistically stationary 
state reads
\begin{equation}
F=\epsilon +\frac{2\eta\nu}{\tau^2} (\langle \mathrm{tr} {\bm \sigma}\rangle -
\mathrm{tr}{\bm 1}) +\alpha \langle |{\bm u}|^2 \rangle
\label{eq:4}
\end{equation}
where $\epsilon=\nu\langle |{\bm \nabla}{\bm u}|^2 \rangle$ is 
the viscous dissipation and $F$ is the average energy input, 
which is flow-independent for a Gaussian, $\delta$-correlated 
random forcing ${\bm f}$.
To obtain eq.~(\ref{eq:4}) we multiply eq.~(\ref{eq:1}) by ${\bm u}$,
add to it the trace of eq.~(\ref{eq:2}) times $\eta \nu/\tau$,
and average over space and time. 
Since in two dimensions kinetic energy flows towards large scales,
it is mainly drained by friction, and
viscous dissipation is vanishingly small in the limit of
very large Reynolds numbers \cite{BCMV02}.
Neglecting $\epsilon$ and observing that 
in the Newtonian case ($\eta=0$) the balance (\ref{eq:4}) 
yields $F=\alpha \langle |{\bm u}|^2 \rangle_{N}$,
we obtain 
\begin{equation}
\langle |{\bm u}|^2 \rangle = 
\langle |{\bm u}|^2 \rangle_{N} - 
\frac{2\eta\nu}{\alpha\tau^2} (\langle \mathrm{tr} {\bm \sigma}\rangle
-\mathrm{tr} {\bm 1})\;.
\label{eq:5}
\end{equation}
Since, as a consequence of incompressibility and chaoticity of the 
flow, it can be shown from eq.~(\ref{eq:2}) that 
$\mathrm{tr} {\bm \sigma} \ge \mathrm{tr} {\bm 1}$,
we finally have
$\langle |{\bm u}|^2 \rangle \le 
\langle |{\bm u}|^2 \rangle_{N}$,
in agreement with numerical results.
This simple energy balance argument can be generalized to nonlinear
elastic models. As viscosity tends to zero,
the average polymer elongation increases so as to compensate for the 
factor $\nu$ in eq.~(\ref{eq:5}), resulting in a finite effect
also in the infinite $Re$ limit. 
Since energy is essentially dissipated by linear friction, the depletion
of $\langle |{\bm u}|^2 \rangle$ entails immediately the reduction
of energy dissipation. The main difference between two-dimensional
``friction reduction'' and three-dimensional drag reduction
resides in the lengthscales involved in the energy drain --
large scales in 2D {\em vs} small scales in 3D.

\begin{figure}[h]
\centerline{\hspace{-0.5cm}
\includegraphics[draft=false,scale=0.7]{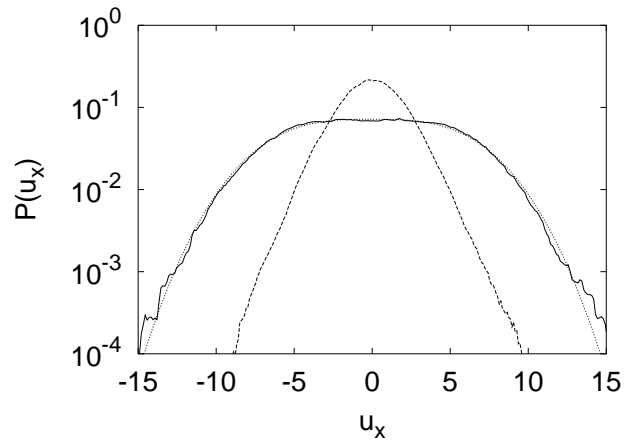} }
\caption{Intermittency of velocity fluctuations induced by polymer additives.
The probability density function
$P(u_x)$ of the velocity component $u_x$ for the Newtonian (continuous line)
and for the viscoelastic case with strong feedback (dashed line). 
Same parameters as in Fig.~\ref{fig:2}. 
Also shown the distribution $\Gamma(2/3) 3^{3/2} \exp(-c|u_x|^3)/(4\pi c)$ 
with $c=2.1\cdot 10^{-3}$ (dotted line).}
\label{fig:4} 
\end{figure}

The effect of polymer additives cannot be merely represented by 
a rescaling of velocity fluctuations by a given factor.
In Fig.~\ref{fig:4} we show the probability distribution
of a velocity component, $u_x$. The choice of the $x$ direction is
immaterial by virtue of statistical isotropy. In the 
 Newtonian case the distribution is remarkably close 
to the subgaussian density $N\exp(-c|u_x|^3)$ stemming 
from the balance between forcing and nonlinear
terms in the Navier-Stokes equation, in agreement with the prediction by
Falkovich and Lebedev \cite{FL97}. On the contrary, the distribution in the 
viscoelastic case is markedly supergaussian, with 
approximately exponential tails. This strong
intermittency in the velocity dynamics is due to the alternation of  
quiescent low-velocity phases ruled by  polymer feedback
and bursting events where inertial nonlinearities take over. 

\begin{figure}[h]
\centerline{\hspace{-0.5cm}
\includegraphics[draft=false,scale=0.7]{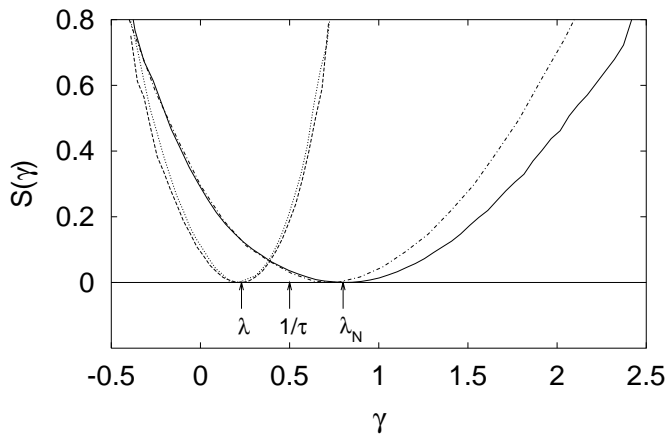} }
\caption{Finite-time Lyapunov exponents decrease in presence of polymers.
The Cram\'er rate function $S(\gamma)$ for the Newtonian 
(continuous line) and for the viscoelastic case 
with strong feedback ($Wi=\lambda_{N} \tau = 1.6$, dashed line). 
Viscosity $\nu=6\cdot 10^{-3}$, 
relaxation time $\tau=2$, $\eta=0.2$ (dashed), $\eta=2$ (dotted).
For sake of completeness, we also show $S(\gamma)$ for a mild feedback case
($Wi=0.4$, $\eta=0.2$, dash-dotted line). 
In the latter case, the Lyapunov exponent
is practically identical to the Newtonian value, and polymers affect
only the right tail of $S(\gamma)$ reducing appreciably the probability
of large stretching events $\gamma \gg \lambda_N$.}
\label{fig:5} 
\end{figure}

Dilute polymers also alter significantly 
the distribution of finite-time Lyapunov exponents $P(\gamma,t)$.
In Fig.~\ref{fig:5} the Cram\'er rate function $S(\gamma) \propto
t^{-1}\ln P(\gamma,t)$
 is shown for the Newtonian 
and for the viscoelastic case. 
Since in the former situation the Lyapunov exponent 
$\lambda_N$ is greater than $1/\tau$, were
the polymers passive all moments of elongation would grow exponentially
fast. However, the feedback can damp stretching so effectively that
after polymer addition $\lambda$ lies below $1/\tau$. This implies
a strong reduction of Lagrangian chaos and a decreased mixing efficiency.  
Moreover, we find that $L_n$ is smaller than $n/\tau$ for all $n$,
a result which guarantees the stationarity of the statistics of 
$\textrm{tr}{\bm \sigma}$ in presence of feedback, while imposing 
a less restrictive condition on 
$S(\gamma)$ than the one proposed in Ref.~\cite{BFL01}.

Finally, we discuss the influence of polymer concentration 
on the properties of the flow. 
As shown in Fig.~\ref{fig:5}, the Cram\'er functions
at two very different values of $\eta$ are practically indistinguishable.
The level of velocity fluctuations (not shown) appears to be
independent of concentration as well. 
This property follows from the viscoelastic equations
assuming that the term $\propto {\bm 1}/\tau$ 
in eq.~(\ref{eq:2}) can be neglected if polymers are substantially stretched.
In that case the dynamics of the conformation tensor 
is invariant under rescaling by a constant
factor, allowing to absorb $\eta$ in the definition of ${\bm \sigma}$,
and making the velocity dynamics independent of concentration.
This observation poses an interesting question: can there be a 
concentration-dependent onset of friction or drag reduction within a 
{\em linear} viscoelastic model ?
Since polymers are increasingly stretched as $\eta \to 0$
there are two alternatives: either 
this is a singular limit, and the passive case is not recovered but
for $\eta$ strictly equal to zero, or the feedback is not uniquely ruled 
by polymer elongation, and there are other relevant mechanisms for
polymer activity, e.g. via the creation of strong 
gradients of the conformation tensor.
The limit of vanishingly small yet finite
$\eta$ is very demanding at the computational level
and its investigation will require further numerical work.

This work has been supported by the EU under the contract
HPRN-CT-2002-00300 and by MIUR-Cofin 2001023848.
Numerical simulations have been performed at CINECA (INFM parallel
computing initiative).


\begin{thebibliography}{99}
\bibitem{note} Equation (\protect{\ref{eq:2}}) is derived 
assuming that the polymer molecules can be 
thought of as linear elastic dumbbells with an end-to-end separation 
${\bm R}$ that obeys to  $\dot{\bm R} = {\bm R}\cdot ({\bm \nabla} {\bm u})
-{\bm R}/\tau + (2 R_0^2/\tau)^{1/2} {\bm \beta}$,
where $R_0$ is the equilibrium size of the polymers and $\tau$ is their
relaxation time.  
The conformation tensor is thus defined as
$\sigma_{ij}=\langle R_i R_j \rangle_{\beta}/ R_0^2$, where $\langle \ldots
\rangle_{\beta}$ denotes the average over the thermal noise ${\bm \beta}$.
\end{thebibliography}
\end{document}